\documentclass[%
 reprint,
amsmath,amssymb,
aps,
prl,
]{revtex4-2}

\usepackage{graphicx}
\usepackage{bm}
\usepackage{braket}
\usepackage{amsthm}

\usepackage{times}
\usepackage{newtxmath}
\usepackage{soul} 
\usepackage{hyperref}
\usepackage[dvipsnames, x11names]{xcolor}
\hypersetup{
  colorlinks,
  breaklinks=true,  citecolor = DeepSkyBlue4, linkcolor = MidnightBlue, urlcolor=MidnightBlue, 
}

\usepackage{orcidlink}
\usepackage[normalem]{ulem}
\begin{document}

\title{
Quantifying two-mode entanglement of bosonic Gaussian states from their full counting statistics
}

\author{Victor Gondret\,\orcidlink{0009-0005-8468-161X}}
\email{victor.gondret@institutoptique.fr}
\author{Clothilde Lamirault\,\orcidlink{0009-0001-6468-2181}}
\author{Rui Dias\,\orcidlink{0009-0004-4158-7693}}
\author{Charlie Leprince\,\orcidlink{0009-0002-5490-6767}}
\author{Christoph I. Westbrook\,\orcidlink{0000-0002-6490-0468}}
\author{David Clément\,\orcidlink{0000-0003-1451-0610}}
\author{Denis Boiron\,\orcidlink{0000-0002-2719-5931}}

\affiliation{
  Université Paris-Saclay, Institut d'Optique Graduate School, CNRS, Laboratoire Charles Fabry, 91127, Palaiseau, France.
}

\begin{abstract}
	\noindent
We study the entanglement properties of two-mode bosonic Gaussian states based on their multi-mode counting statistics. 
We exploit the idea that measuring high-order correlations of particle numbers can reveal entanglement without making any assumptions about the coherence of the fields.
We show that the two- and four-body number correlations are sufficient to fully characterize the entanglement of two-mode bosonic Gaussian states for which each mode exhibits a thermal distribution.
In addition, we derive an entanglement witness based on two-body correlations alone.
Our findings are of great importance because it becomes possible to reveal entanglement in a series of recent experiments.
\\

\end{abstract}
\maketitle

Entanglement of bosonic Gaussian states is a key ingredient of continuous-variable quantum information theory \cite{weedbrook.gaussian.2012} that has found many applications in Quantum Optics. 
It is also expected to emerge in a variety of other situations described by Hamiltonians that are quadratic in field operators, ranging from Bogoliubov pairing in superfluids \cite{bogoliubov1947theory}, Hawking emission at the black hole horizon \cite{hawking.1974.black_hole} to the dynamical Casimir effect \cite{moore.dce.1970}.
Existing criteria which detect such entanglement are often expressed in terms of inequalities involving the expectation values, products, and variances of \textit{field} operators \cite{duan.2000.inseparability, simon.peres.horodecki.2000, hillery.entanglement.2006}.
Measuring these quantities requires some variant of homodyne or heterodyne detection in which the states of interest are made to interfere with coherent local oscillators \cite{vogel.1989.determination,ferris.2008.detection}.
Although these methods have been successfully implemented experimentally in some specific configurations \cite{ou.1992.epr, gross_atomic_2011, peise_satisfying_2015}, they can be difficult to apply in general.
This is the case for example, for massive particles entangled in their external degrees of freedom \cite{bergschneider_experimental_2019, greve_entanglement.2022,athreya2025bellcorrelationsmomentumentangledpairs},
when the spatial mode to be probed does not match that of the local oscillator, or simply when the number of modes in the system is large \cite{Christ_2011}.
Characterization of quantum states would benefit from finding entanglement criteria that do not require such homodyne techniques. 

\begin{figure}
    \centering
    \includegraphics[width=\linewidth]{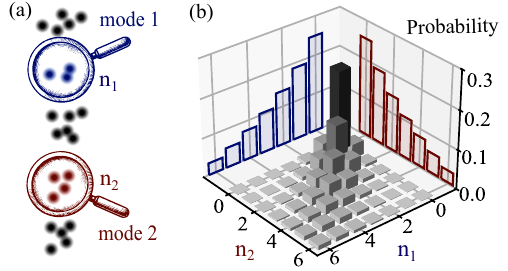}
    \caption{
    (a) Two modes in a system are defined (shown by the lens here) and the number of particles $n_1$ and $n_2$  in each mode is counted.
    (b) These counts can be histogrammed to obtain the full counting statistics of each mode. In addition, the joint probability distribution is also accessible.
    This information yields the $n$-body correlation functions   of particle number operators of any order.
    }
    \label{fig:scheme}
\end{figure}

On the other hand, particle number resolved detection methods \cite{zadeh.2021.superconducting,Ott2016} allow quantum states to be characterized via their multi-mode counting statistics \cite{dall.ideal.2013, schweigler_experimental_2017, bergschneider_experimental_2019, perrier.thermal.2019, herce.full.2023,lantz.2023.multiphoton, allemand.prx.2024,  ketterle.2024.insitu, krishnaswamy.2024.experimental, finger.2024.spin,burenkov.2017.fullstat}, as sketched in Fig. \ref{fig:scheme}.
The information extracted about a bosonic quantum state yields any order of normally ordered correlation functions of the \textit{particle number} operators \textit{i.e.} $\braket{:\hat{n}_1^{\alpha_1}\hat{n}^{\alpha_2}_2:}$ where the $\alpha_i$'s are integers, $\hat{n}_i=\hat{a}_i^\dagger\hat{a}_i$ and $\hat{a}_i$ is the annihilation operator of mode $i$.
These observables thus involve the same number of annihilation and creation operators of a given mode.
As such, the multi-mode correlations of the \textit{field}, such as $\langle\hat{a}_1 \hat{a}_2^\dagger\rangle$ or $\langle\hat{a}_1 \hat{a}_2\rangle$, cannot be directly obtained from the counting statistics. 
In general, this lack of information hinders the ability to probe entanglement properties using existing criteria.

By contrast, the particle number correlation function of Gaussian states can be written in terms of one and two-field correlation functions [see Eqs. (\ref{eq:g2_def}) and (\ref{eq:g4_def})], 
and, in the context of Quantum Optics, their full counting statistics has recently gained attention as a valuable tool for their characterization~\cite{ fiurasek.squeezing.2004, wenger.pulsed.2004, Sperling.2013.correlation,perina.2017.nonclassicality, burenkov.2017.fullstat, zuo.2016.determination,kumar.2020.optimal, avagyan.2023.multimode}.
In particular, Barasi\'{n}ski \textit{et al.} \cite{barasinski.2023.quantification} have shown that the measurement of particle number correlation functions up to the fourth order can be used to witness entanglement.
In this work, we focus on two-mode Gaussian states for which the single-mode statistics of each mode exhibit a thermal probability distribution. 
Well known in quantum optics, such two-mode Gaussian states have been produced in various cold-atom experiments, ranging from the parametric amplification of non-interacting quasi-particles \cite{jaskula.dce.2012,clark_collective_2017, chen.2021.observation, schemmer.2018} and the Bogoliubov pairing in interacting Bose gases \cite{tenart.observation.2021, Bureik2025}, to
analog Hawking radiation \cite{steinhauer_observation_2016} and the quasi-particle creation in an expanding metric \cite{oberthaler.2024.particles}. 
Thus far, observations of positive two-mode correlations were related to two-mode entanglement only upon \textit{assuming} the absence of coherence between the two modes \cite{steinhauer_observation_2016, tenart.observation.2021, chen.2021.observation}. A similar assumption is made when evaluating entanglement from the time evolution of the density-density correlation function within Bogoliubov theory \cite{robertson.2017.assessing, agullo2024observationentangledpairsbec}.\\

We show how the measurement of higher order correlation functions can assess and quantify entanglement, relying solely on  the two-mode full counting statistics.
We consider two-mode Gaussian states of bosons for which single modes exhibit thermal statistics, a property which can be verified in experiments. 
Elaborating on the positivity of the partial transpose (PPT) criterion \cite{peres.separability.1996} and on the properties of Gaussian states, we introduce an entanglement criterion based on the mode populations and the two- and four-body correlation functions as obtained from measuring the full counting statistics. 
We also derive an entanglement witness based on the two-body correlation function alone.
\\


\textit{Hypothesis} --
In the following, we consider two-mode bosonic Gaussian states  
satisfying  $\braket{\hat{a}_i}= 0 $ and $\braket{\hat{a}_i^2}=0$.
The latter hypothesis is equivalent to requiring that each single mode distribution is thermal \cite{avagyan.2023.multimode}.
Note that ``thermal" refers here to the statistical properties of a mode and does not imply a thermodynamic temperature nor thermal equilibrium.
Thermal single-mode distributions are expected in many situations \cite{finger.2024.spin,jaskula.dce.2012,  clark_collective_2017,chen.2021.observation, schemmer.2018,steinhauer_observation_2016, tenart.observation.2021, Bureik2025, oberthaler.2024.particles} and can be verified experimentally as shown in Refs. \cite{dall.ideal.2013, perrier.thermal.2019, herce.full.2023}.

The normalized two-body correlation function of these zero mean Gaussian states can be expanded using Wick's theorem~\cite{Kardar_2007} or using the characteristic function of the Gaussian state~\cite{barasinski.2023.quantification}
\begin{equation}\label{eq:g2_def}
g^{(2)}_{12}:=\frac{\langle\hat{a}^\dagger_1\hat{a}^\dagger_2\hat{a}_1\hat{a}_2\rangle}{n_1n_2} = 1 + \frac{|\langle\hat{a}_1\hat{a}_2\rangle|^2}{n_1n_2} + \frac{|\langle\hat{a}_1\hat{a}_2^\dagger\rangle|^2}{n_1n_2}.
\end{equation}
Any separable state must satisfy $|\langle\hat{a}_1\hat{a}_2\rangle |^2 \leq n_1n_2$ \cite{hillery.entanglement.2006}. 
Thus, the observation of $g^{(2)}_{12}> 2$ can be considered a signature of entanglement only \textit{under the assumption} that the coherence between the two modes vanishes, \textit{i.e.}, $\braket{\hat{a}_1\hat{a}_2^\dagger}=0$. This is the assumption that we aim to drop by considering higher-order correlations.

Within our hypothesis, the four-body correlation function between the two modes is given by
\begin{equation}\label{eq:g4_def}
    \begin{split}
        g^{(4)}_{12} := \frac{\braket{(\hat{a}_1^\dagger\hat{a}_2^\dagger)^2(\hat{a}_1\hat{a}_2)^2}}{n_1^2n_2^2} = 4&\Biggl[ 1  + \left(g^{(2)}_{12} - 1\right)^2 \\
        + 4\left(g^{(2)}_{12} - 1\right)
        +2 &\frac{|\langle\hat{a}_1\hat{a}_2^\dagger\rangle|^2|\langle\hat{a}_1\hat{a}_2\rangle|^2}{n_1^2 n_2^2} \Biggr].
    \end{split}
\end{equation}
Importantly, while $g^{(2)}_{12}$ involves the quadratic sum of $ |\langle\hat{a}_1\hat{a}_2^\dagger\rangle| $ and $ |\langle\hat{a}_1\hat{a}_2\rangle| $, the normalized four-body correlation function $g^{(4)}_{12}$ includes their product. Therefore, Eqs. (\ref{eq:g2_def}) and (\ref{eq:g4_def}) constitute a symmetric system to determine the values of $|\langle\hat{a}_1\hat{a}_2\rangle|$ and $|\langle\hat{a}_1\hat{a}_2^\dagger\rangle|$ and can characterize entanglement without further assumptions.\\


\textit{Entanglement criterion -}  
A two-mode Gaussian state with thermal single-mode statistics is entangled if and only if the smallest value of the set of the symplectic eigenvalues of its covariance matrix and its partial transpose is strictly smaller than one. 
This value, $\lambda_-$, is defined in Eq.~(\ref{eq:smallest_symplectic}) and solely depends on the measurement of the populations $n_1$ and $n_2$ and of the two- and four-body correlation functions $g^{(2)}_{12}$ and $g^{(4)}_{12}$.

\begin{proof}
Solving the system consisting of Eqs. (\ref{eq:g2_def}) and (\ref{eq:g4_def}) for $|\langle\hat{a}_1\hat{a}_2\rangle|$ and $|\langle\hat{a}_1\hat{a}_2^\dagger\rangle|$ yields two solutions $\beta_\pm$ given by 
\begin{equation}\label{eq:beta_pm}
     \beta_\pm^2 = n_1n_2(g_{12}^{(2)} - 1)\frac{ 1\pm \sqrt{1 - \theta }}{2},
\end{equation}
where $\theta$ is defined by
\begin{equation}\label{eq:def_theta}
\theta =\frac{g^{(4)}_{12}+12 - 16g^{(2)}_{12} -4\left(g^{(2)}_{12}-1\right)^2}{2\left(g^{(2)}_{12}-1\right)^2}.
\end{equation}
The quantities $\beta_\pm$ in Eq.~(\ref{eq:beta_pm}) correspond to the values of  $|\langle\hat{a}_1\hat{a}_2\rangle|$ and $|\langle\hat{a}_1\hat{a}_2^\dagger\rangle|$ but, because the system is symmetric, the distinction between the two solutions is not \textit{a priori} possible.
However, the covariance matrix of a Gaussian state must satisfy a so-called \textit{bona fide} condition which imposes an inequality on its symplectic eigenvalues \cite{serafini.quantum.2017}.
These constraints identify the physical solutions.
In appendix \ref{sec:gaussian_state} we show that, within our hypotheses:
\begin{itemize}
    \item[\textit{(i)}] the symplectic eigenvalues of a two-mode Gaussian state only depend on $n_1$, $n_2$, $|\langle\hat{a}_1\hat{a}_2\rangle|$ and $|\langle\hat{a}_1\hat{a}_2^\dagger\rangle|$ (see Eq.~(\ref{eq:symplectic_eigenvalues})) and must be greater or equal to one to describe a physical state,
    
    \item[\textit{(ii)}] the partial transpose operation simply interchanges the roles of $|\langle\hat{a}_1\hat{a}_2\rangle|$ and $|\langle\hat{a}_1\hat{a}_2^\dagger\rangle|$. According to the PPT criterion \cite{simon.peres.horodecki.2000}, the state is entangled if and only if its partial transpose is not a \textit{bona fide} Gaussian state, \textit{i.e.}, if at least one of the symplectic eigenvalues of the partially transposed state is smaller than one. 
    These eigenvalues are found as in \textit{(i)}, with $|\langle\hat{a}_1\hat{a}_2\rangle|$ and $|\langle\hat{a}_1\hat{a}_2^\dagger\rangle|$ interchanged.
\end{itemize}
Based on \textit{(i)} and \textit{(ii)}, we conclude that the measurement of $\beta_\pm$ allows us to compute the set of the symplectic eigenvalues of the state and of its partial transpose. 
In particular, $\lambda_{-}$, the smallest symplectic eigenvalue of this set, is given by \cite{serafini.symplectic.2004}
\begin{equation}\label{eq:smallest_symplectic}
     2\lambda_{-}^2 = \Delta - \sqrt{\Delta^2-4\text{det}\boldsymbol{\sigma}}
\end{equation}
where 
\begin{equation}\label{eq:detsigma}
    \begin{split}
    \text{det}\boldsymbol{\sigma}=&\, 16  \left( \beta_+^2 -\beta_-^2   \right)^2+ (1 + 2 n_1)^2(1 + 2 n_2)^2\\
    & \,- 8 \left(\beta_+^2 +\beta_-^2\right) (1 + 2 n_1)(1 + 2 n_2)
\end{split}
\end{equation}
is the determinant of the covariance matrix $\boldsymbol{\sigma}$ and 
\begin{equation}
    \Delta = (2n_1+1)^2 + (2n_2+1)^2 + 
    8(\beta_+^2-\beta_-^2).
\end{equation}
If $\lambda_-\geq 1$, all the symplectic eigenvalues of the state and its partial transpose are greater than one. Because of \textit{(ii)}, the state is therefore separable.\newline
If  $\lambda_-<1$, $\lambda_-$ cannot be a symplectic eigenvalue of the state, because of \textit{(i)}. Thus, $\lambda_-$ must be a symplectic eigenvalue of the partially transposed state, which means that the state is entangled from \textit{(ii)}.
In this case, we can also conclude $|\langle\hat{a}_1\hat{a}_2\rangle|=\beta_+$ and $|\langle\hat{a}_1\hat{a}_2^\dagger\rangle|=\beta_-$.
\end{proof}


\textit{Logarithmic negativity -} 
The logarithmic negativity is useful to quantify entanglement \cite{vidal.2002.computable}.
For instance, it increases with the squeezing strength \cite{isoard.2021.bipartite,brady.2022.symplectic}, and provides an upper bound for the number of distillable Bell states \cite{vidal.2002.computable}. 
It can be computed from the smallest symplectic eigenvalue of the partially transposed state.
Within our assumptions (see above), the logarithmic negativity is
\begin{equation}\label{eq:logneg}
   \text{LN}= \text{max}(-\log_2( \lambda_-), 0).
\end{equation}
When the state is separable ($\lambda_- \geq 1$), the logarithmic negativity is zero while it is strictly positive when the state is entangled, ($\lambda_-<1$). \newline
\begin{figure}
    \centering
    \includegraphics[width=1\linewidth]{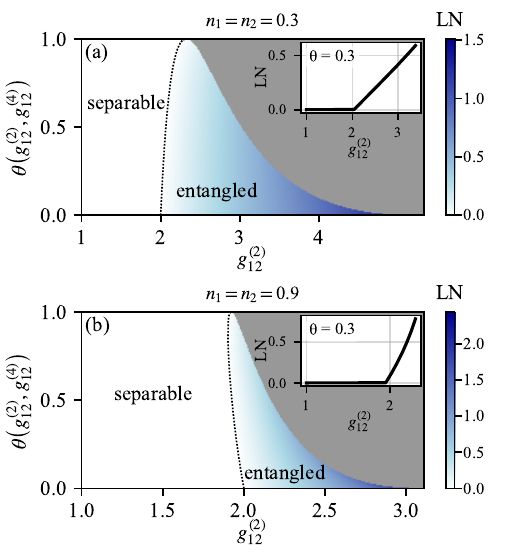}
     \caption{
     Two-mode entanglement in the $(g^{(2)}_{12}, \theta)$ plane for equal mean  populations  $n_1=n_2=0.3$ (a) and 0.9 (b). The parameter $\theta$ is defined in Eq.~(\ref{eq:def_theta}) and varies linearly with $g^{(4)}_{12}$. The grey region corresponds to unphysical states or states not fulfilling our hypotheses and the color-scale to the logarithmic negativity: the bluer, the more entangled. The entanglement border (dotted black line) splits the plane between separable (left) and entangled states (right). The inset shows logatithmic negativity as a function of $g^{(2)}_{12}$ for $\theta=0.3$.
}
    \label{fig:g2g4plane}
\end{figure}
The logarithmic negativity is plotted in Fig. \ref{fig:g2g4plane} as a function of $\theta$ and $g^{(2)}_{12}$ for two populations, assuming that $n_1=n_2$. 
On the left of the black dotted curve lie separable states while entangled states lie on the right. 
The shape of this entanglement border depends on the population. 
On the $x$-axis, $\theta=0$ which gives $\beta_-=0$. 
The entangled states then correspond to two-mode squeezed thermal states for which $\braket{\hat{a}_1\hat{a}_2^\dagger}=0$.
The two-mode squeezed vacuum state is the one having the maximal value of $g^{(2)}_{12}$.
The parameter $\theta$ should lie in the interval $[0,1]$ since $\beta^2_\pm$ are by definition positive, see Eq.~(\ref{eq:beta_pm}). 
This condition imposes bounds on the possible values of $(g^{(2)}_{12},g^{(4)}_{12})$. 
This is also the case for the gray region in which states do not satisfy condition \textit{(i)}.
The case of non-equal populations $n_1 \neq n_2$ is qualitatively similar, but will not be discussed further.

\textit{Detector efficiency - } On the one hand, $g^{(2)}_{12}, g^{(4)}_{12}$ and $\theta$ are normalized quantities that do not depend on the detection efficiency $\eta$. 
On the other hand, the populations ($n_1$, $n_2$) depend linearly on $\eta$ \cite{martin.2023.comparing}. 
As a result, one must scale the measured mean populations by $1/\eta$ to use the results of Fig. \ref{fig:g2g4plane} with a imperfect detector ($\eta <1$).
Consider a state with ($\theta=0.5, g^{(2)}_{12}=2.03$) and a measured mean population $n_1=n_2=0.3$:  with a unit detection efficiency $\eta=1$ one concludes from Fig. \ref{fig:g2g4plane} that the state is separable; in contrast, if $\eta= 1/3$, one concludes that the state is entangled.\\

\textit{Two-body witnesses -} To witness entanglement or separability, the measurement of the four-body correlation function is not necessary. 
From Fig. \ref{fig:g2g4plane}(a), it is evident that $g^{(2)}_{12}>2.3$ (entanglement border for $\theta=1$) guarantees entanglement, whereas Fig. \ref{fig:g2g4plane}(b) reveals that $g^{(2)}_{12}>2$ is enough. 
The entanglement threshold thus depends on the curvature of the entanglement border which in turn depends on the population of the state. 
In Appendix \ref{sec:g2_appendix}, we demonstrate that a state is entangled if $g^{(2)}_{12}>g^{(2)}_{E}$ where 
\begin{equation}\label{eq:g2_non_sep}
    \begin{split}
      \text{if } n_1n_2 < 1/2,\, \, \, \, &   g^{(2)}_{E} =2 +\frac{1/2 -n_1n_2}{2n_1n_2 + n_1 + n_2 +1/2},\\
      \text{if } n_1n_2 \geq 1/2,\, \, \, \, &   g^{(2)}_{E} =2. \\
    \end{split}
\end{equation}
\begin{figure}
    \centering
    \includegraphics[width=\linewidth]{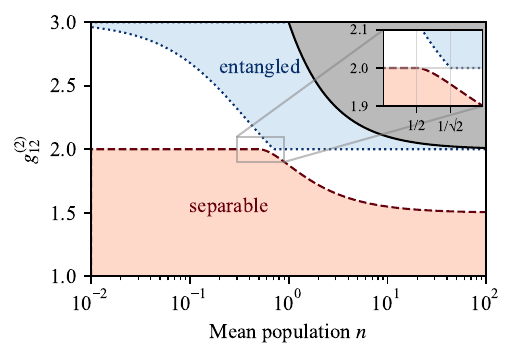}
    \caption{Entanglement $g_E^{(2)}$ (dotted blue) and separability $g^{(2)}_S$ (dashed red) thresholds for the two-body correlation function as a function of the mean population of the state $n_1=n_2=n$, in log scale. The four-body correlation function is needed to discriminate entanglement in the white region located in between the separable and entangled regions. The unphysical gray region is above the line $2+1/n$ which is the value of $g^{(2)}_{12}$ for a two-mode squeezed vacuum state with population $n$. }
    \label{fig:g2witness}
\end{figure}
The entanglement threshold $g^{(2)}_E$ is plotted in Fig. \ref{fig:g2witness} as a blue dotted line in the case of equal populations, $n_1=n_2=n$. 
When the mean populations are sufficiently high ($n_1n_2\ge 1/2$), a value of $g^{(2)}_{12}>2$ is sufficient to witness entanglement. 
When the mean population is smaller, the entanglement threshold $g^{(2)}_E$ exceeds 2. 
It reaches 3 in the limit of vanishing populations arising from the general inequality $|\langle\hat{a}_1\hat{a}_2^\dagger\rangle|^2<n_1n_2$ (see Appendix \ref{sec:g2_appendix}). \newline
Likewise, the quantum states located on the left of the entanglement border in Fig. \ref{fig:g2g4plane} are separable, allowing the derivation of a separability threshold based solely on the value of $g^{(2)}_{12}$. We demonstrate in Appendix  \ref{sec:g2_appendix} that, if $g^{(2)}_{12}\leq g^{(2)}_{S}$, the state is separable where 
\begin{equation}\label{eq:g2_sep}
    \begin{split}
      \text{if } n_1n_2 \leq 1/4,\, \, \, \, &   g^{(2)}_{S} =2 ,\\
      \text{if } n_1n_2 > 1/4,\, \, \, \, &   g^{(2)}_{S} =2 -\frac{(1-4n_1n_2)^2}{8n_1n_2(1+2n_1)(1+2n_2)}. \\
    \end{split}
\end{equation}
This separability threshold, depicted as a red dashed line in Fig. \ref{fig:g2witness}, equals 2 when the populations fulfill $n_1n_2<1/4$ and approaches 1.5 at large values of the mean populations.

These two thresholds act as entanglement and separability witnesses and not as criteria:
they cannot certify the (non-)separability of the state in the white region of Fig. \ref{fig:g2witness}, where a measurement of the four-body correlation function is required. 

This is illustrated by Fig.~\ref{fig:g2thetafixed} in the appendix, which shows how the thresholds in Fig.~\ref{fig:g2witness} effectively become criteria when the four-body correlation is known.
The grey area in Fig.~\ref{fig:g2witness} is bounded by the line $g^{(2)}_{12} = 2 + 1/n$, which corresponds to the value for a two-mode squeezed vacuum state with mean population $n$. This boundary defines the minimal unphysical region, obtained for $\theta = 0$. 
As $\theta$ increases, the unphysical region expands towards lower population and lower $g^{(2)}_{12}$, as illustrated in Fig.~\ref{fig:g2thetafixed} in the appendix. 
\\


\textit{Discussion} - 
In this work, we focus on entanglement between \textit{modes}, following the non-separability definition of Werner \cite{werner.quantum.1989}.
This is in contrast to the so-called \textit{particle} entanglement that emerges from the indistinguishability of particles~\cite{morris.2020.entanglement},
Defining the ratio $\mathcal{C_S}= g^{(2)}_{12}/\left(g^{(2)}_{1}g^{(2)}_{2}\right)^{1/2}$ where $g^{(2)}_{i} = \braket{(\hat{a}_i^\dagger)^2\hat{a}_i^2}/n_i^2$, Ref. \cite{wasak.2014.cauchy} shows that the violation of the classical Cauchy-Schwarz inequality $\mathcal{C_S} \leq 1$ is a sufficient condition for particle entanglement.
For the single-mode thermal statistics we consider here, one has $g^{(2)}_{1}=g^{(2)}_{2}=2$ and $\mathcal{C_S}=g^{(2)}_{12}/2$. 
Mode entanglement ($g^{(2)}_{12}>g_{E}^{(2)}\geq 2$) therefore necessarily implies particle entanglement. 
Our witness for mode entanglement is thus more stringent than the one for particle entanglement.

Our results apply to Gaussian states for which single-mode distributions are thermal, which is not always true.
For example, recent work \cite{agullo.2022.quantum,delhom.2024.entanglement} has suggested using a single-mode squeezed vacuum state at an analog black-hole horizon to boost the effect of the two-mode squeezer.
The output state is expected to exhibit super-thermal statistics with $g^{(2)}_i>2$.
In this scenario, the symplectic spectrum depends on the relative phase between the modes, and the four-body correlation function contains additional terms with respect to those of Eq.~(\ref{eq:g4_def}).
However following Ref. \cite{barasinski.2023.quantification}, an entanglement witness can still be obtained by measuring the two-, three- and four-body correlation functions.
An interesting question for follow-up studies is whether an entanglement criterion could be established in such situations as well, by extending our approach to higher orders of the counting statistics (\textit{e.g.} using $g^{(6)}_{12}$).

Finally, the entanglement criterion introduced using the values of $g^{(2)}_{12}$ and $g^{(4)}_{12}$ does not require the measurement of non-commuting observables.
While this may appear puzzling at first, we emphasize that this is the consequence of the assumption that the states are Gaussian states.
Proving unambiguously that a state is indeed Gaussian would require the measurement of its quadratures \cite{finke.observation.2016,finazzi.2014.entangled},  \textit{i.e.} the implementation of a homodyne scheme.

\textit{Conclusion -}
We have shown that measuring both the two- and four-body correlation functions is necessary to unambiguously determine the entanglement of a two-mode Gaussian state with single-mode thermal statistics.
Still, the two-body correlation function alone provides an entanglement witness that can be used to analyze some already published experimental data.
Based on our results, recent work on four-wave mixing induced by the collision of Bose-Einstein condensates reported by Ref.
\cite{hodgman_solving_2017} and to the Bogoliubov quantum depletion observed by Ref. \cite{tenart.observation.2021} are sufficient to claim entanglement, without any additional assumptions on the field coherence $\langle\hat{a}_1\hat{a}_2^\dagger\rangle$. \\

\textit{Data availability --} The code which generates the figures that support the findings of this article are openly available at~\cite{gondret.2025.code}.

\section*{Acknowledgments}
\noindent We acknowledge fruitful discussions with Amaury Micheli and Scott Robertson, and thank Jarom\'\i{}r Fiur\'a\v{s}ek for providing us with insightful references.
We also acknowledge  Léa Camier and Amaury Micheli for their careful reading and feedback on the manuscript.
V.G. thanks the organizers of the \textit{Analog Gravity in 2023} summer school, and Antony Brady for his lecture on Gaussian states.
The research leading to these results has received funding from QuantERA Grant No. ANR-22-QUA2-000801 (MENTA), ANR Grant No. 20-CE-47-0001-01 (COSQUA), France 2030 programs of the French National Research Agency (Grant numbers ANR-22-PETQ-0004), Région Ile-de-France in the framework of the DIM SIRTEQ program, and the Quantum Saclay program, supported by the state under France 2030 (reference ANR-21-CMAQ-0002).
R.D. acknowledges a PhD grant with reference 2024.03181.BD from the Portuguese Foundation for Science and Technology (FCT).


\appendix
\section{Gaussian state formalism in a nutshell}\label{sec:gaussian_state}
We consider the two-mode Gaussian reduced density matrix of a quantum state.
This (reduced) state is thus fully characterized by its first and second moments: the mean vector (null based on our hypothesis) and the covariance matrix~\cite{serafini.quantum.2017}.
We work in the creation and annihilation operator basis $\boldsymbol{\hat{r}}=(\hat{a}_1, \hat{a}_1^\dagger, \hat{a}_2, \hat{a}_2^\dagger)^\intercal$ and define the (complex hermitian) covariance matrix by $\boldsymbol{\sigma}_{ij}= \braket{\boldsymbol{\hat{r}}_i \boldsymbol{\hat{r}}_j^{\dagger} +\boldsymbol{\hat{r}}_j^{\dagger}\boldsymbol{\hat{r}}_i }$.
Within our hypothesis, this matrix is given by \cite{serafini.quantum.2017}
\begin{equation}\label{eq:covariance_matrix_block}
    \boldsymbol{\sigma}=\begin{pmatrix} 
\boldsymbol{A} & \boldsymbol{C}\\
\boldsymbol{C}^\dagger & \boldsymbol{B}
\end{pmatrix},
\end{equation}
where $ \boldsymbol{A}=(2n_1+1)\mathbb{I}_2$ and  $ \boldsymbol{B}=(2n_2+1)\mathbb{I}_2$ are the covariance matrices of the respective reduced states composed of modes $1$ and $2$ and $\mathbb{I}_2$ is the identity matrix. $\braket{\hat{a}_j^2}\neq 0$ would imply the presence of non-zero off-diagonal terms in these matrices.
Sub-matrix $\boldsymbol{C}$ encodes the correlation between the two modes and is given by
\begin{equation}
    \boldsymbol{C} = 2 \begin{pmatrix} 
    \langle\hat{a}_1\hat{a}_2^\dagger\rangle & 
\langle\hat{a}_1\hat{a}_2\rangle \\
\langle \hat{a}_1\hat{a}_2\rangle^* & \langle\hat{a}_1\hat{a}_2^\dagger\rangle^*
\end{pmatrix}. \label{eq:matrixC}
\end{equation}
Because the covariance matrix is a positive semi-definite matrix, it can be diagonalized through symplectic transformations \textit{i.e.} transformations that preserve the canonical commutation relations \cite{serafini.quantum.2017}.
The symplectic eigenvalues  $\nu_\pm$  of this matrix are given by \cite{serafini.symplectic.2004}
\begin{equation}\label{eq:symplectic_eigenvalues}
    2\nu_\pm^2 = \Gamma \pm \sqrt{\Gamma^2-4\text{det}\boldsymbol{\sigma}}
\end{equation}
where $\Gamma = \text{det}\boldsymbol{A} +\text{det}\boldsymbol{B}+2\text{det}\boldsymbol{C}$. From Eq.~(\ref{eq:matrixC}), $\text{det}\boldsymbol{C}$ does not depend on the relative phase between $\braket{\hat{a}_1\hat{a}_2^\dagger} $ and $\langle\hat{a}_1\hat{a}_2\rangle$.
The value of $\text{det}\boldsymbol{\sigma}$ is given by 
Eq.~(\ref{eq:detsigma}) of the main text, where $\beta_\pm$ should be replaced by the values of $|\langle\hat{a}_1\hat{a}_2\rangle|$ and $|\langle\hat{a}_1\hat{a}_2^\dagger\rangle|$.
We see that the values of $\nu_\pm$ do not depend on the relative phase between $\braket{\hat{a}_1\hat{a}_2^\dagger} $ and $\langle\hat{a}_1\hat{a}_2\rangle$, which is a consequence of the $\braket{\hat{a}_j^2}=0$ hypothesis.

With our convention, the \textit{bona fide} condition on the covariance matrix is equivalent to $\nu_\pm\geq 1$ \cite{brady.2022.symplectic}.
Positivity of the partial transpose (PT) state is a necessary and sufficient condition for the separability of the state \cite{simon.peres.horodecki.2000}. 
Here, the partial transpose operation on mode 2 transforms the covariance matrix  as
\begin{equation}
    \boldsymbol{A} \xrightarrow{\text{\sc pt}} \boldsymbol{A},\quad
    \boldsymbol{B} \xrightarrow{\text{\sc pt}} \boldsymbol{\sigma}_x\boldsymbol{B\sigma}_x,\quad
    \boldsymbol{C} \xrightarrow{\text{\sc pt}} \boldsymbol{C\sigma}_x
\end{equation}
where $\boldsymbol{\sigma}_x=\begin{pmatrix}0 &1\\1&0\end{pmatrix}$.
Within our hypothesis, the partial transpose only exchanges the roles of $ \braket{\hat{a}_1\hat{a}_2^\dagger}$ and $ \langle\hat{a}_1\hat{a}_2\rangle$.

The symplectic eigenvalues of the partial transpose covariance matrix $\tilde{\nu}_\pm$ are thus given by Eq.~(\ref{eq:symplectic_eigenvalues}) where $\Gamma$ is replaced by $\tilde{\Gamma} = \text{det}\boldsymbol{A} +\text{det}\boldsymbol{B}-2\text{det}\boldsymbol{C}$. 
The determinant of the covariance matrix is preserved under partial transpose operation \cite{serafini.symplectic.2004}, which also explains why Eq.~(\ref{eq:detsigma}) is invariant under the exchange of $\beta_\pm$.
Thus we conclude that the two-mode Gaussian state is entangled iff $\tilde{\nu}_-<1$.
The expression $\lambda_-=\text{min}(\nu_-, \tilde{\nu}_-)$ defined in Eq.~(\ref{eq:smallest_symplectic}) in the main text is given by Eq.~(\ref{eq:symplectic_eigenvalues}) replacing $\Gamma$ by  $\Delta= \text{det}\boldsymbol{A} +\text{det}\boldsymbol{B}+2|\text{det}\boldsymbol{C}|$.
As discussed by Simon \cite{simon.peres.horodecki.2000}, an entangled state must have $\text{det}\boldsymbol{C}<0$ to respect the \textit{bona fide} condition.
Thus, when the state is entangled, inequality $|\langle\hat{a}_1\hat{a}_2\rangle|>|\langle\hat{a}_1\hat{a}_2^\dagger\rangle|$ must hold and because $\beta_+\geq\beta_-$, we can identify $|\langle\hat{a}_1\hat{a}_2\rangle|=\beta_+$ and $|\langle\hat{a}_1\hat{a}_2^\dagger\rangle|=\beta_-$.

The grey region in Fig. \ref{fig:g2g4plane} corresponds to states which do not satisfy the \textit{bona fide} condition.
For these states, the two solutions $(|\langle\hat{a}_1\hat{a}_2\rangle|,|\langle\hat{a}_1\hat{a}_2^\dagger\rangle|)=(\beta_\pm, \beta_\mp)$ are unphysical: both $\nu_-$ and $\tilde{\nu}_-$ are smaller than 1. 
Thus, $\lambda_-'=\text{max}(\nu_-, \tilde{\nu}_-)$ is smaller than 1 where $\lambda_-'$  is given by Eq.~(\ref{eq:smallest_symplectic}) where $\Delta$ should be replaced by 
\begin{equation}
   \begin{split}
    \Delta' &=  \text{det}\boldsymbol{A} +\text{det}\boldsymbol{B}-2|\text{det}\boldsymbol{C}|\\
    &=  (2n_1+1)^2 + (2n_2+1)^2 - 
    8(\beta_+^2-\beta_-^2).
\end{split}
\end{equation}

\textit{Detection efficiency - } The effect of non-unit efficiency $\eta$ can be modeled by a pure loss channel that mixes each mode with the vacuum through a beam-splitter with a transmitivity $\eta$.
Such an operation does not alter the Gaussianity of the state and transforms the covariance matrix $\boldsymbol{\sigma}$ according to $\eta \boldsymbol{\sigma} +(1-\eta)\mathbb{I}_4$ \cite{martin.2023.comparing}.
With our parametrization, the values $n_1$, $n_2$, $\braket{\hat{a}_1\hat{a}_2^\dagger}$ and $\langle\hat{a}_1\hat{a}_2\rangle$ are divided by $\eta$. 
In the main text, the values of $\beta_\pm$ are proportional to the populations and should be rescaled accordingly when taking into account the non-unit efficiency.

\section{Entanglement and separability thresholds for $\mathbf{g^{(2)}}$}\label{sec:g2_appendix}

As we saw in the last section, a two-mode Gaussian state must satisfy the \textit{bona fide} condition $\nu_\pm\geq 1$ and such a state is entangled if and only if $\tilde{\nu}_-<1$ . As a result, a two-mode Gaussian state is entangled if only if $\text{min}(\nu_-,\tilde{\nu}_-)<1$, or equivalently $\mathcal{P}_- <0$ where \cite{simon.peres.horodecki.2000, serafini.symplectic.2004,brady.2022.symplectic}
\begin{equation}
    \mathcal{P}_- = 1+\text{det}\boldsymbol{\sigma}-\text{det}\boldsymbol{A} -\text{det}\boldsymbol{B}-2|\text{det}\boldsymbol{C}|.
\end{equation}
This condition being necessary and sufficient, such a state is separable if and only if $\mathcal{P}_-\geq 0$.
As the second order correlation function gives access to  $| \langle\hat{a}_1 \hat{a}_2\rangle|^2+ | \langle\hat{a}_1 \hat{a}_2^\dagger\rangle|^2$, it is natural to introduce  
\begin{equation}\label{eq:delta_def}
    \delta=\big| | \langle\hat{a}_1 \hat{a}_2\rangle|^2 - | \langle\hat{a}_1 \hat{a}_2^\dagger\rangle|^2 \big|/n_1n_2
\end{equation} 
so that $\mathcal{P}_-$ is given by 
\begin{equation}\label{eq:p_minus_delta2}
    \begin{split}
\mathcal{P}_- = 16 n_1n_2\biggl[ & (1 + n_1)(1 + n_2)(2 -g^{(2)}_{12}  ) \\
 & +\left( \frac{1}{2}- n_1n_2\right) \left(g^{(2)}_{12}  -1\right) \\
 &+  \delta\left(n_1n_2\delta  - \frac{1}{2}\right) \biggr].
\end{split}
\end{equation}
The value of $\delta$ cannot take arbitrarily large or small values.
In particular, since the density matrix of the state is positive semi-definite, the following Cauchy-Schwarz inequality for any operators \(\hat{\phi}\) and \(\hat{\psi}\) must be satisfied \cite{adamek.2013.dissipative}
\begin{equation}\label{eq:cauchy_schwarz}
|\langle\hat{\phi}^\dagger \hat{\psi}\rangle|^2 \leq \braket{\hat{\phi}^\dagger\hat{\phi}}\braket{\hat{\psi}^\dagger\hat{\psi}}.
\end{equation}
This inequality cannot be violated and that the so-called violation of the classical Cauchy-Schwarz inequality refers to a normally ordered version of Eq.~(\ref{eq:cauchy_schwarz}) \cite{adamek.2013.dissipative}.
This inequality yields
\begin{equation}\label{eq:cs_inequality}
\begin{split}
    | \langle\hat{a}_1 \hat{a}_2\rangle|^2 \leq  & n_1n_2 + n_1, \\
  | \braket{\hat{a}_2 \hat{a}_1}|^2  \leq & n_1n_2 + n_2,\\
   |\langle\hat{a}_1 \hat{a}_2^\dagger\rangle|^2  \leq & n_1n_2.\\
\end{split}
\end{equation}
Assuming that $n_1\leq n_2$ without loss of generality, we see that $\delta$ must lie within $[0,1+1/n_2]$.\\

\textit{Lemma}: If $\delta>1$,
the state is entangled.
\begin{proof}
If $\delta>1$, then at least one of the two terms $|\langle\hat{a}_1\hat{a}_2\rangle|^2/n_1n_2$ and $|\langle\hat{a}_1\hat{a}_2^\dagger\rangle|^2/n_1n_2$ is strictly greater than 1. 
Since the Cauchy-Schwarz inequality (\ref{eq:cs_inequality}) gives $ |\langle\hat{a}_1 \hat{a}_2^\dagger\rangle|^2  \leq n_1n_2 $, then we must have $|\langle\hat{a}_1\hat{a}_2\rangle|^2>n_1n_2$ which implies entanglement from Ref.~\cite{hillery.entanglement.2006}.
\end{proof}

\textit{Entanglement threshold:} A two-mode Gaussian state for which each mode is thermal and which satisfies $g^{(2)}_{12}>g^{(2)}_E$ is entangled. $g_E^{(2)}$ depends on the population and is given in Eq.~(\ref{eq:g2_non_sep}).
\begin{proof}
$\mathcal{P}_-$ is a strictly decreasing function of $g^{(2)}_{12}$, therefore if $\mathcal{P}_-(g^{(2)}_E, \delta)\leq 0$ for the allowed values of $\delta$, then any state satisfying $g^{(2)}_{12}>g^{(2)}_E$ is entangled. 
We therefore look for $g^{(2)}_E$ which is the highest value of the normalized two-body correlation function for a separable state \textit{i.e.} the ``last one". 
Fixing the populations $n_1$ and $n_2$, $g^{(2)}_E$ is found solving
\begin{equation}
    \text{Max}_\delta\left\{\mathcal{P}_-(n_1, n_2, g^{(2)}_E, \delta)\right\}=0.
\end{equation}
We now seek an upper bound for the expression $\delta(n_1n_2\delta -1/2)$  in Eq.~(\ref{eq:p_minus_delta2}) and subsequently solve $\mathcal{P}_-=0$. 
From the above lemma, any state with $\delta>1$ is entangled ($\mathcal{P}_-<0$). 
As we aim to find the ``last'' \textit{separable} state ($\mathcal{P}_-=0$), we restrict our analysis to $\delta\in [0,1]$.
We are left to distinguish two cases.
\begin{itemize}
    \item If $n_1n_2<1/2$, the last term of equation (\ref{eq:p_minus_delta2}) is always negative, and an upper bound for this term is obtained for $\delta=0$. Solving $\mathcal{P_-}=0$ leads to $g^{(2)}_{E}$ given in Eq.~(\ref{eq:g2_non_sep}). 
    \item If $n_1n_2>1/2$, an upper bound is found for $\delta=1$. In this scenario, the second and third terms of Eq.~(\ref{eq:p_minus_delta2}) simplify, and solving $\mathcal{P}_-=0$ yields $g^{(2)}_{E}=2$.
\end{itemize}
\end{proof}
\begin{figure}
    \centering
    \includegraphics[width=\linewidth]{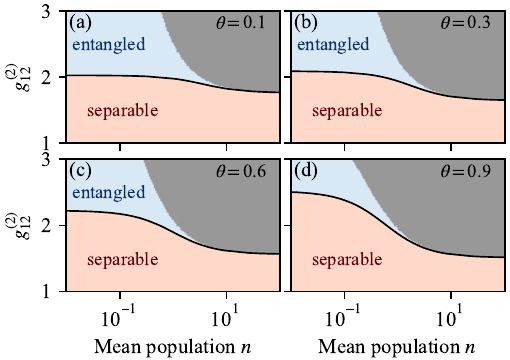}
\caption{Entanglement regions in the $(n,g^{(2)}_{12})$ plane, for a given value of $\theta$ in each panel. 
Compared to the entanglement witness shown in Fig. \ref{fig:g2witness}, there is no unknown entanglement region.}    
\label{fig:g2thetafixed}
\end{figure}
In Fig. \ref{fig:g2thetafixed}, we show the entanglement criterion in the $(n, g^{(2)}_{12})$ plane for four values of $\theta$.
The entanglement threshold $g_E^{(2)}$ can  be graphically understood from the behavior of the entanglement border (black line). 
At fixed population $n$, $g^{(2)}_E$ corresponds to corresponds to the maximum of the values of $g^{(2)}_{12}$ that belong to the entanglement border with respect to $\theta$. 
For small populations,  we see in Fig. \ref{fig:g2thetafixed} that, at fixed population, the values of $g_{12}^{(2)}$ on the entanglement border are larger than 2.
The maximum value is given for $\theta=1$,  
which implies $\beta_+=\beta_-$ thus $| \langle\hat{a}_1 \hat{a}_2\rangle| = | \langle\hat{a}_1 \hat{a}_2^\dagger\rangle| $ and $\delta=0$.
On the other hand, for large population, the values of $g^{(2)}_{12}$ of the entanglement border lie below 2, and the maximum  is given for $\theta=0$ for which $g^{(2)}_{12}=2$.
\\

\textit{Separability threshold:} A two-mode Gaussian state for which each mode is thermal and which satisfies $g^{(2)}_{12}\leq g^{(2)}_S$ is separable. $g_S^{(2)}$ depends on the population and is given in Eq.~(\ref{eq:g2_sep}).
\begin{proof}
$\mathcal{P}_-$ is a strictly decreasing function of $g^{(2)}_{12}$, therefore if $\mathcal{P}_-(g^{(2)}_S, \delta)\geq 0$ for all allowed values of $\delta$, then any state satisfying $g^{(2)}_{12}\leq g^{(2)}_S$ is separable.
From Eq.~(\ref{eq:p_minus_delta2}), the minimum of the $\mathcal{P}_-(\delta)$ polynomial is reached for $\delta = 1/4n_1n_2$. 
But here again, $\delta$ cannot take arbitrarily large values: it must lie within $[0,1]$, as we show below.\newline
When the coherence is null, we know that separable states satisfy $g^{(2)}_{12}\leq 2$ thus this specific case means that $g^{(2)}_{S}$ satisfies $g^{(2)}_{S}\leq 2$.
However, $\delta>1$ implies that $g^{(2)}_{12}>2$.
As we require $g^{(2)}_{12}\leq g^{(2)}_{S}\leq 2$, we conclude that $\delta\leq 1$ and thus $\delta\in [0,1]$. 
We are left to distinguish two cases.
\begin{itemize}
    \item If $n_1n_2 \leq 1/4$, the minimum of the $\mathcal{P}_-$ polynomial on $[0,1]$ is 1, and solving $\mathcal{P}_-(\delta=1)=0$  yields $g^{(2)}_{S}=2$.
    \item If $n_1n_2>1/4$, the minimum of $\mathcal{P}_-$ is reached for $\delta_{\text{min}} = 1/4n_1n_2$. Solving $\mathcal{P}_-(\delta_{\text{min}})=0$ yields $g^{(2)}_S$ given in Eq.~(\ref{eq:g2_non_sep}).
\end{itemize}
\end{proof}

\bibliography{biblio}

\end{document}